\documentclass{ws-mplb}

\begin{document}

\markboth{V. Thanh NGO, D. Tien HOANG, H. T. DIEP}{Phase Transition in Fully Frustrated Lattice}

\catchline{}{}{}{}{}
%
%%%%%%%%%%%%%%%%%%%%%%%%%%%%%%%%%%%%%%%%%%%%%%%%%%%%%%%%%%%%%%%%%%%%

\title{PHASE TRANSITION IN HEISENBERG FULLY FRUSTRATED SIMPLE CUBIC LATTICE}

\author{\footnotesize V. Thanh NGO\footnote{nvthanh@iop.vast.ac.vn} \footnote{Also at Division of Fusion and Convergence of Mathematical
Sciences, National Institute for Mathematical Sciences, Daejeon, Republic of
Korea.} and D. Tien HOANG\footnote{hdtien@iop.vast.ac.vn}}

\address{Institute of Physics, P.O. Box 429,   Bo Ho, Hanoi 10000,
Vietnam\\}

\author{H. T. DIEP\footnote{diep@u-cergy.fr} \footnote{Corresponding author}}

\address{Laboratoire de Physique Th\'eorique et Mod\'elisation,
Universit\'e de Cergy-Pontoise,\\ CNRS UMR 8089\\
2, Avenue Adolphe Chauvin, 95302 Cergy-Pontoise Cedex, France.}

\maketitle

\begin{history}
\received{(Day Month Year)}
\revised{(Day Month Year)}
\end{history}

\begin{abstract}
The phase transition in frustrated spin systems is a fascinated subject in statistical physics. We show the result obtained by the Wang-Landau flat histogram Monte Carlo simulation
on the phase transition in the fully frustrated simple cubic lattice with the Heisenberg spin model.  The degeneracy of the ground state of this system is infinite with two continuous parameters.  We find a clear first-order transition in contradiction with previous studies which have shown a second-order transition with unusual critical properties.   The robustness of our calculations allows us to conclude this issue putting an end to the 20-year long uncertainty.
\end{abstract}


\section{Introduction}

During the last 30 years, intensive investigations have been carried out to study the effect of the frustration
in spin systems.  The frustration is known to be the origin of many unusual properties such as large ground state (GS)
degeneracy, successive phase transitions, partially disordered phase, reentrance and disorder lines.
Frustrated systems still constitute at present a challenge for theoretical physics.\cite{Diep2005}

One of the most studied aspects is the nature of the phase transition in frustrated spin systems.  Exact methods have been devised to solve with mathematical elegance many problems in two dimensions.\cite{Diep-Giacomini,Diep-Debauche91} Numerical simulations and various approximations have been used to study three-dimensional frustrated cases. In particular, numerical simulations which did not need huge memory and long calculations for simple non frustrated systems require now new devices, new algorithms to improve convergence in frustrated systems.\cite{Ngo08,Diep2008}

Frustrated systems are very unstable due to the competition between different kinds of interaction. However, they have no disorder and therefore can be exactly solved in two dimensions.\cite{Diep-Giacomini,Diep-Debauche91} In three dimensions, it is not the case: several systems are not well understood.\cite{Diep2005}   Let us recall the definition of a frustrated system.
When a spin cannot fully satisfy energetically all the interactions with its neighbors, it is
"frustrated".   This occurs when the interactions are in competition with
each other or when the lattice geometry does not allow to satisfy all interaction bonds simultaneously. A well-known example is the stacked triangular antiferromagnet (STA) with interaction between nearest-neighbors (NN). This system with Ising,\cite{Diep93}  XY and Heisenberg spins\cite{Delamotte2004,Loison} have been intensively studied since 1987,\cite{kawamura87,kawamura88,azaria90,Loison94,Boub,Dobry,antonenko,Loison2000} but only recently that the 20-year controversy comes to an end.\cite{Ngo08,Diep2008,tissier00b,tissier00,tissier01,itakura03,Peles,Kanki,Bekhechi,Zelli}
For details, see for example the review by Delamotte
et al.\cite{Delamotte2004}

In this work, we study another fully frustrated model called fully frustrated simple cubic lattice (FFSCL) shown  in Fig. \ref{fig:SCFF}. A detailed description of the model will be presented in section \ref{Model}. The nature of the phase transition in the classical XY spin model has been recently investigated.\cite{Ngo2010}  It was shown that it is a first-order transition putting an end to a 20-year long controversial issue. In this paper, we study the case of Heisenberg spin model.

Section \ref{Model} is devoted to the
description of the model and  some technical details of the Wang-Landau
(WL) methods as applied in the present paper.  Section \ref{Res} shows our
results.  Concluding remarks are given in section \ref{Concl}.

\section{Model and Wang-Landau Method}\label{Model}

\begin{figure}
\centerline{\epsfig{file=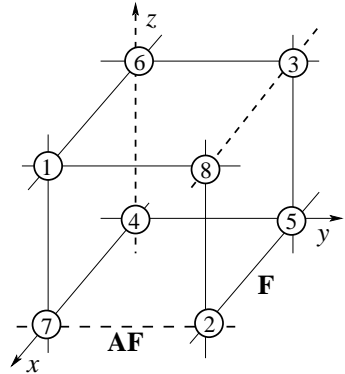,width=1.8in}} \caption{Fully frustrated simple cubic lattice.  Discontinued (continued) lines denote antiferromagnetic (ferromagnetic) bonds.} \label{fig:SCFF}
\end{figure}

We consider the fully frustrated simple cubic lattice (FFSCL) shown  in Fig. \ref{fig:SCFF}.  The Hamiltonian is given by
\begin{equation}
{\cal H} = -\sum_{(i,j)} J_{ij}\mathbf{S}_i.\mathbf{S}_j,
\end{equation}
where $\mathbf{S}_i$ is the classical Heisengerg spin of magnitude $S=1$ at the lattice site $i$, $\sum_{(i,j)}$ is made
over the NN spin pairs $\mathbf{S}_i$ and $\mathbf{S}_j$ with  interaction $J_{ij}$.
We take $J_{ij}=-J$ ($J>0$) for antiferromagnetic bonds indicated  by discontinued lines in Fig. \ref{fig:SCFF}, and $J_{ij}=J$ for ferromagnetic bonds.  This is the three dimensional counterpart of the Villain's model.\cite{Villain,Berge,Lee91,Diep98}

 Let us recall some results on the present model. The GS are given by the following three independent relations which determine the relative orientation of every spin pair\cite{Diep85c,Diep85b}

\begin{eqnarray}
\mathbf {S}_2\cdot \mathbf {S}_3+\mathbf {S}_3\cdot \mathbf {S}_4+\mathbf {S}_2\cdot \mathbf {S}_4&=&0\\
-\mathbf {S}_1\cdot \mathbf {S}_3+\mathbf {S}_3\cdot \mathbf {S}_4+\mathbf {S}_1\cdot \mathbf {S}_4&=&0\\
\mathbf {S}_1\cdot \mathbf {S}_2+\mathbf {S}_2\cdot \mathbf {S}_4-\mathbf {S}_1\cdot \mathbf {S}_4&=&0
\end{eqnarray}
 For the XY model, there are 12 non collinear planar configurations.\cite{Diep85c,Ngo2010}  For the classical Heisenberg model,  the GS degeneracy is infinite with two free parameters.\cite{Diep85c}  The reader is referred to those papers for the details of the GS calculation.

%\begin{figure}
%\centerline{\epsfig{file=GS.eps,width=2.0in}}
%\caption{One the the 12 ground-state configurations of the fully frustrated simple cubic lattice.
%The numbers indicate the spins at the sites defined in Fig. 1, $\beta=\pi /4$,
%$\alpha=\arccos(\frac{1+\sqrt{2}}{\sqrt {6}})$.} \label{fig:GS}
%\end{figure}

This model has been studied by standard Monte Carlo (MC) simulation with small lattice sizes, short runs
and poor statistics more than 20 years ago.\cite{Diep85b}  The result has shown a second order transition with unusual critical exponents.   MC technique and computer capacity at that time did not allow us to conclude the matter with certainty.

Wang and Landau\cite{WL1} have recently proposed a MC algorithm  which allowed to study classical statistical models with difficultly accessed microscopic states. In particular, it permits to detect  with efficiency weak first-order transitions.\cite{Ngo08,Diep2008,Ngo2010} The algorithm uses a random walk in energy space in order to obtained an accurate estimate for the density of states $g(E)$ which is defined as the number of spin configurations for any given $E$. This method is based on the fact that a flat energy histogram $H(E)$ is produced if the probability for the transition to a state of energy $E$ is proportional to $g(E)^{-1}$.

We summarize how this algorithm is implied here. At the beginning of the simulation, the density of states (DOS) is set equal to one for all possible energies, $g(E)=1$.
We begin a random walk in energy space $(E)$ by choosing a site randomly and flipping its spin with a probability
proportional to the inverse of the temporary density of states (DOS). In general, if $E$ and $E'$ are the energies before and after a spin is flipped, the transition probability from $E$ to $E'$ is
\begin{equation}
p(E\rightarrow E')=\min\left[g(E)/g(E'),1\right].
\label{eq:wlprob}
\end{equation}
The details of the WL method as applied to our spin models have been given in our recent papers.\cite{Ngo08,Diep2008,Ngo2010}  We shall not repeat it here.  We just emphasize the following point.
We consider here an energy range of interest\cite{Schulz,Malakis}
$(E_{\min},E_{\max})$. We divide this energy range to $R$ subintervals, the minimum energy of the $i-th$ subinterval is $E^i_{\min}$ ($i=1,2,...,R$), and the maximum  is $E^i_{\max}=E^{i+1}_{\min}+2\Delta E$,
where $\Delta E$ can be chosen large enough for a smooth boundary between two subintervals. The WL
algorithm is used to calculate the relative DOS of each subinterval $(E^i_{\min},E^i_{\max})$ with a flatness criterion $x\%=95\%$.
Note that we reject a spin flip and do not update $g(E)$ and the energy histogram $H(E)$ of
the current energy level $E$ if the spin-flip trial would result in an energy outside the energy segment.
The DOS of the whole range is obtained by joining the DOS of each
subinterval $(E^i_{\min}+\Delta E,E^i_{\max}-\Delta E)$.

The thermodynamic quantities\cite{WL1,brown} can be evaluated by
\begin{eqnarray}
\langle E^n\rangle &=&\frac{1}{Z}\sum_E E^n g(E)\exp(-E/k_BT)\\
C_v&=&\frac{\langle E^2\rangle-\langle E\rangle^2}{k_BT^2}\\
\langle M^n\rangle &=&\frac{1}{Z}\sum_E M^n g(E)\exp(-E/k_BT)\\
\chi&=&\frac{\langle M^2\rangle-\langle M\rangle^2}{k_BT}
\end{eqnarray}
where $Z$ is the partition function defined by
%\begin{equation}
$Z =\sum_E g(E)\exp(-E/k_BT)$.
%\label{eq:partfunc}
%\end{equation}
The canonical distribution at a temperature $T$ can be calculated simply by
%\begin{equation}
$P(E,T) =\frac{1}{Z}g(E)\exp(-E/k_BT)$.
%\label{eq:pe}
%\end{equation}

\section{Results}\label{Res}

The following system sizes have been used in our simulations $N\times N \times N$ where
$N$ varies from 24 up to 90. At $N=90$, as seen below, the transition shows a definite answer to the problem studied here.  Periodic
boundary conditions are used in the three directions.  $J=1$ is
taken as the unit of energy in the following.

\begin{figure}
\centerline{\epsfig{file=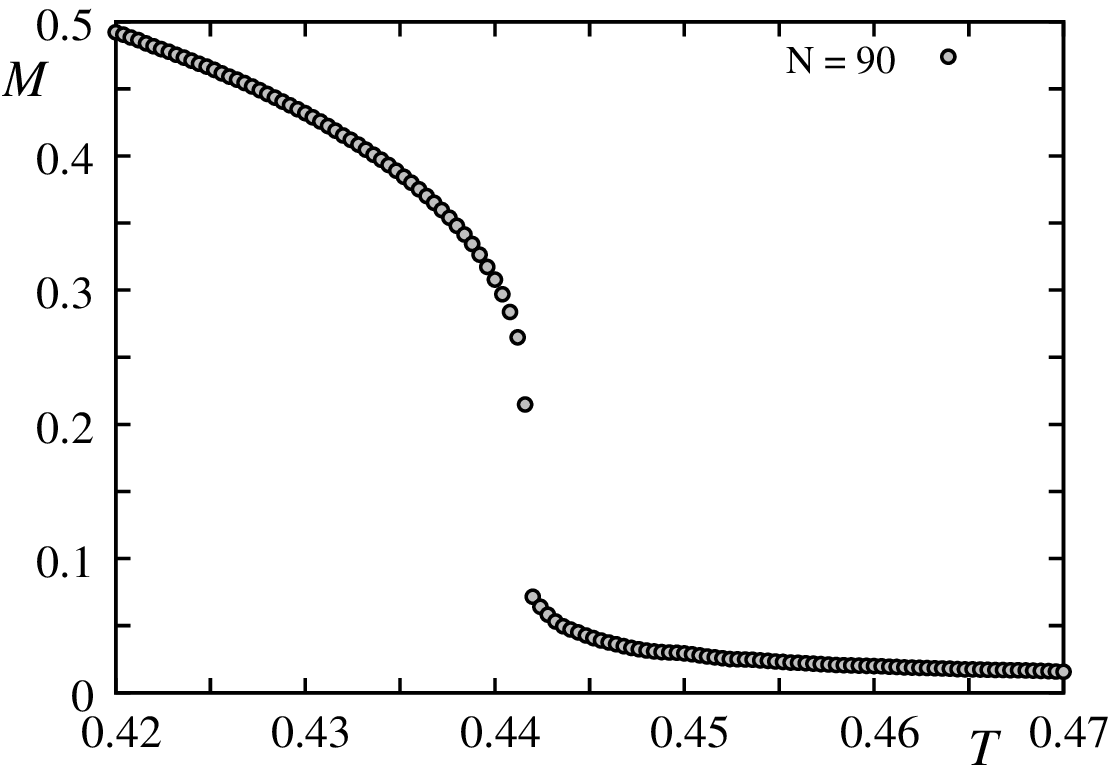,width=2.1in}}
\centerline{\epsfig{file=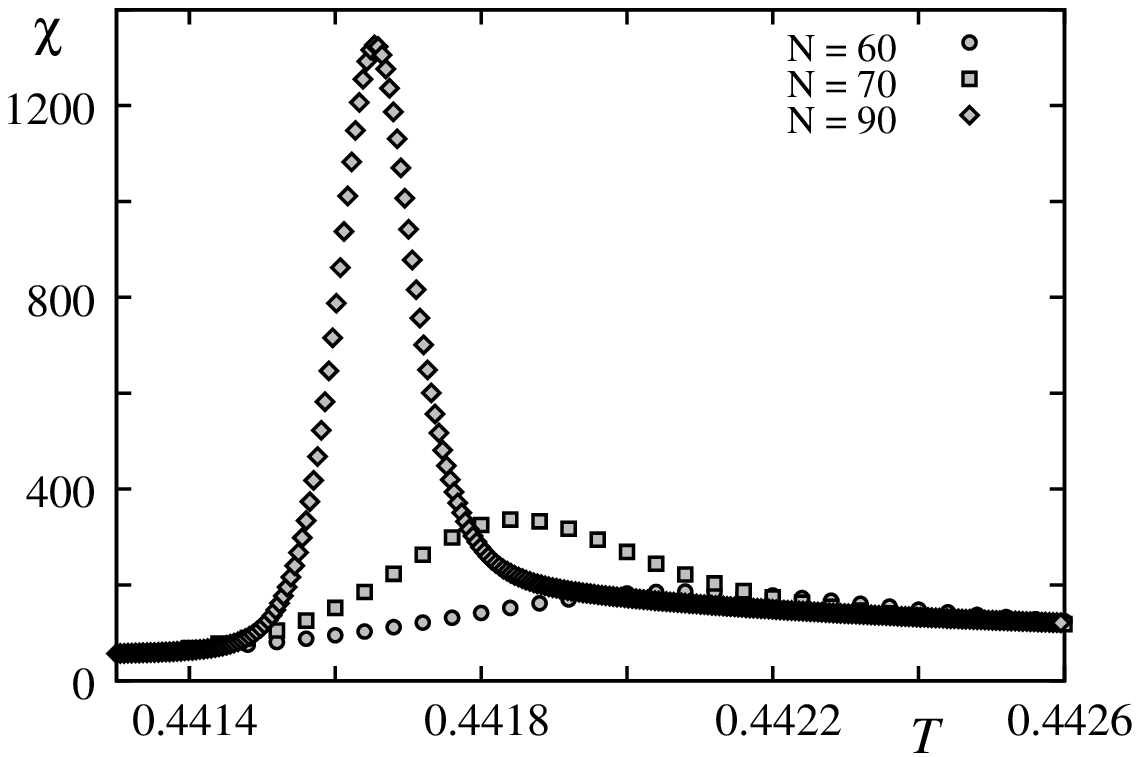,width=2.1in}}
\caption{Magnetization (upper curve) for $N=90$ and susceptibility (lower curve) for $N=60$, 70, 90,
 versus $T$.} \label{fig:MX}
\end{figure}

\begin{figure}
\centerline{\epsfig{file=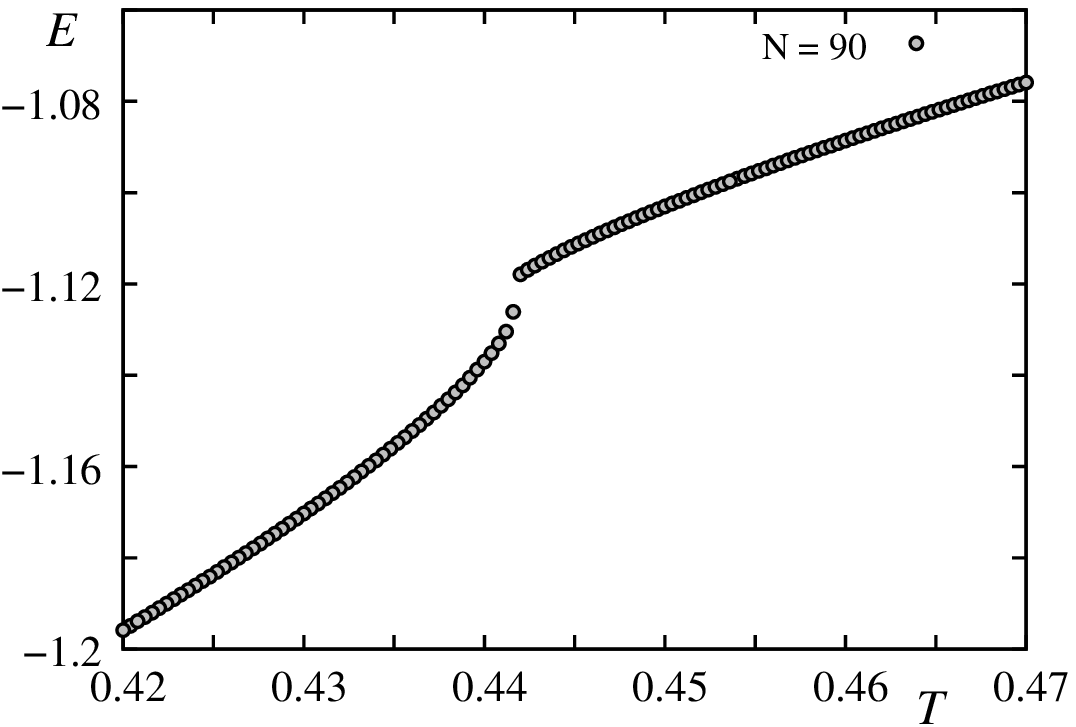,width=2.1in}}
\centerline{\epsfig{file=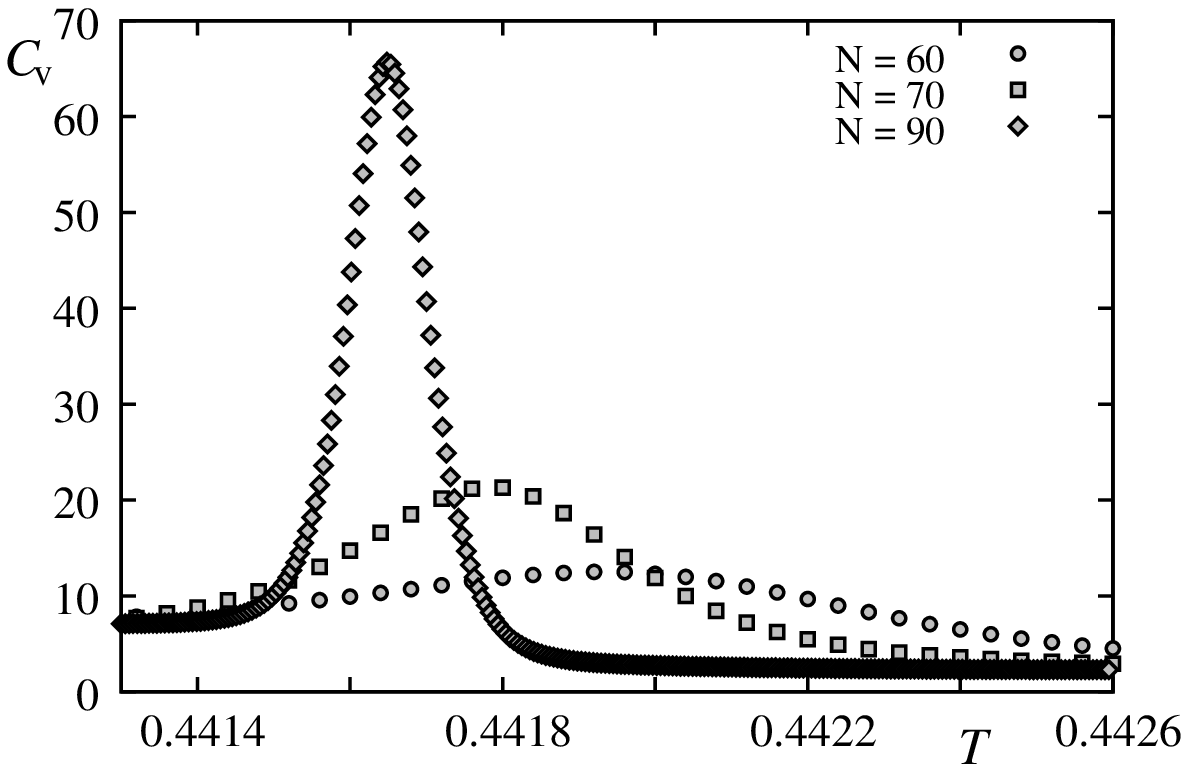,width=2.1in}}
\caption{Energy per spin for $N=90$ (upper curve) and specific heat per spin  (lower curve) for $N=60$, 70, 90,
 versus $T$.} \label{fig:EC}
\end{figure}

\begin{figure}
\centerline{\epsfig{file=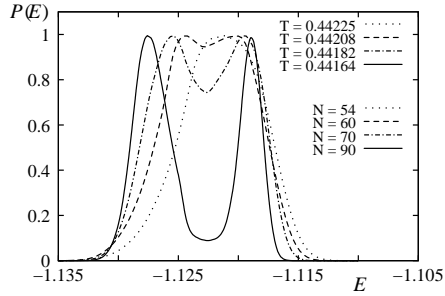,width=2.3in}} \caption{Energy
histogram for several sizes $N=54$, 60, 70, 90 at $T_c$ indicated on the
figure.}\label{fig:PE}
\end{figure}

Figure \ref{fig:MX} shows, as functions of $T$,  the magnetization for $N=90$, and the susceptibility for $N=60$, 70 and 90. These curves show a sharp transition but they do not allow us to conclude about a first-order character.  The same observation is for
the energy per spin and the specific heat shown in Fig. \ref{fig:EC}  for $N=90$. In this situation where there is a possibility of very weak first-order transitions, the WL method is very useful because it allows us to sample rarely accessed microscopic states by establishing a flat DOS.
The energy histograms obtained by WL technique for four representative sizes $N=54$, 60, 70 and 90 are shown in Fig. \ref{fig:PE}. As seen, for $N=54$, the energy histogram, though unusually broad, shows a single peak indicating a continuous energy at the transition. The double-peak histogram starts only from $N=60$ and the dip between the two maxima becomes deeper with increasing size, as observed at $N=70$ and 90.  We note
that the distance between the two peaks, i. e.
the latent heat, increases with increasing size and reaches $\simeq 0.0085$ for $N=90$.  This is rather small compared
with the value $\simeq 0.03$ for $N=48$ in the XY case.\cite{Ngo2010}

The double-peak structure is a clear signature of a first-order transition. It indicates a discontinuity in energy at the transition and gives the latent heat.  Without an efficient MC method, weak first-order transition cannot be easily detected.
We give here the values of $T_c$ for a few sizes:  $T_c =
0.44225\pm0.00010$, $0.44208\pm 0.00010$, $0.44182\pm 0.00010$ and $0.44164\pm 0.00010$ for $N=54$, 60, 70  and 90,
respectively.

%\begin{figure}
%\centerline{\epsfig{file=STA36PE.eps,width=2.8in}} \caption{Energy
%histogram for $N=36$.} \label{fig:STA36PE}
%\end{figure}

%\begin{figure}
%\centerline{\epsfig{file=STA36GE.eps,width=2.8in}}
%\caption{Density of state for $N=36$.} \label{fig:STA36GE}
%\end{figure}

%\begin{figure}
%\centerline{\epsfig{file=STAH3CV.eps,width=2.5in}} \caption{Specific heat
% versus $T$  for  $N=96, 120, 150$.}
%\label{fig:STAH3CV}
%\end{figure}

\section{Concluding Remarks}\label{Concl}

To conclude, let us emphasize by using the powerful WL flat histogram technique, we have studied the phase transition in the Heisenberg fully frustrated simple cubic lattice. In weak first-order transitions, the
technique is very efficient because it helps to overcome extremely long
transition time between energy valleys.   We found that the transition is clearly of first-order at large lattice sizes in contradiction of early studies using standard MC algorithm and much smaller sizes.\cite{Diep85b}

The result presented here will serve as a testing ground for theoretical methods such as the nonperturbative renormalization group which has recently succeeded in clarifying the nature of the transition in the much debated STA with vector spins.\cite{Delamotte2004}  We note that some other three-dimensional Heisenberg frustrated systems such as the FCC,\cite{Diep89} HCP\cite{Diep92} and helimagnetic\cite{Diep89a} antiferromagnets show also a first-order transition in MC simulations. It would be interesting to check if it is a general rule or not.

\section*{Acknowledgements}
One of us (VTN) would like to thank  the University of Cergy-Pontoise for a financial support during the course of this work. He is grateful to Nafosted of Vietnam National Foundation
for Science and Technology Development, for support (Grant No. 103.02.57.09). He also thanks the NIMS (National Institute for Mathematical
Sciences, Korea) for hospitality and financial support.

\end{document}